\documentclass{article}

\usepackage{tabulary,graphicx,times,caption,fancyhdr,amsfonts,amssymb,amsbsy,latexsym,amsmath,amsthm}
\usepackage{graphicx}
\usepackage{float}
\usepackage[ruled, linesnumbered,noend,lined,boxed]{algorithm2e}
\usepackage{extraplaceins}
\usepackage{epstopdf}
\usepackage[utf8]{inputenc}
\usepackage{url,multirow,morefloats,floatflt,cancel,tfrupee,textcomp,colortbl,xcolor,pifont}
\usepackage[nointegrals]{wasysym}
\makeatletter

\usepackage{ifxetex}
\ifxetex\else
  \usepackage{dblfloatfix}
\fi

\@ifundefined{subparagraph}{
\def\subparagraph{\@startsection{paragraph}{5}{2\parindent}{0ex plus 0.1ex minus 0.1ex}%
{0ex}{\normalfont\small\itshape}}%
}{}

\def\URL#1#2{\@ifundefined{href}{#2}{\href{#1}{#2}}}

\def\UrlOrds{\do\*\do\-\do\~\do\'\do\"\do\-}%
\g@addto@macro{\UrlBreaks}{\UrlOrds}

\makeatother

\usepackage[paperheight=11in,paperwidth=8.3in,margin=2.5cm,headsep=.7cm,top=2.5cm]{geometry}
\usepackage[T1]{fontenc}

\renewenvironment{abstract}
	{\trivlist\item[]\leftskip0pt\par\vskip4pt\noindent
  	\textbf{\abstractname}\mbox{\null}\\}
	{\par\noindent\endtrivlist}

\def\keywords#1{\par\medskip\par\noindent\textbf{Keywords}: #1\par}

 \date{} \emergencystretch 8pt

\makeatletter
\def\author#1{\gdef\@author{\hskip-\tabcolsep%
	\parbox{\textwidth}{\raggedright\bfseries#1\\[1pc]}}}
\def\address[#1]#2{\g@addto@macro\@author{\\\hskip-\tabcolsep\parbox{\textwidth}{\raggedright%
	\normalsize\normalfont\textsuperscript{#1}#2}}}
\let\addresslink\textsuperscript
\def\correspondence#1{\g@addto@macro\@author{\\\hskip-\tabcolsep\parbox{\textwidth}{\raggedright%
	\vspace*{10pt}\normalsize\normalfont~\\#1~\\[12pt]}}}
\def\email#1{\g@addto@macro\@author{\\\hskip-\tabcolsep\parbox{\textwidth}{\raggedright%
	\normalsize\normalfont Emails: #1}}}

\def\title#1{\gdef\@title{\vspace*{-30pt}%
	\raggedright\textbf{\@journaltitle}~\\%
  \raggedright\bfseries\ifx\@articleType\@empty\vspace*{20pt}\else%
  \vspace*{20pt}\@articleType\vspace*{20pt}\\\fi#1}}
\let\@journaltitle\@empty \def\journaltitle#1{\gdef\@journaltitle{{\normalfont\itshape#1}}}
\let\@articleType\@empty \def\articletype#1{\gdef\@articleType{{\normalfont\itshape#1}}}

\let\@runningHead\@empty \def\RunningHead#1{\gdef\@runningHead{{\normalfont #1}}}

\usepackage{fancyhdr}
\fancypagestyle{headings}{\fancyhf{}
  \fancyhead[R]{\itshape\@runningHead}
  \fancyfoot[C]{\thepage}}
\fancypagestyle{plain}{%
	\fancyhf{}\fancyhead[R]{}
  \fancyfoot[C]{\thepage}}
\makeatother

\usepackage[%
	numbers,sort&compress%
  ]{natbib}

\usepackage{float,xcolor}

\begin{document}

\title{Computing a many-to-many matching with demands and capacities between two sets using the Hungarian algorithm}

\author{%
		Fatemeh Rajabi-Alni\addresslink{1}, and
  	Alireza Bagheri\addresslink{1}}
		
\address[1]{Computer Engineering Department, Amirkabir University of Technology (Tehran Polytechnic), Tehran, Iran}


\email{f.rajabialni@aut.ac.ir (F. Rajabi-Alni), ar\_bagheri@aut.ac.ir (A. Bagheri)}%


\maketitle

\begin{abstract}
Given two sets $A=\{a_1,a_2,\dots,a_s\}$ and $\{b_1,b_2,\dots,b_t\}$, a \textit{many-to-many matching with demands and capacities} (MMDC) between $A$ and $B$ matches each element $a_i \in A$ to at least $\alpha_i$ and at most ${\alpha '}_i$ elements in $B$, and each element $b_j \in B$ to at least $\beta_j$ and at most ${\beta '}_j$ elements in $A$ for all $1 \leq i \leq s$ and $1 \leq j \leq t$. In this paper, we present an algorithm for finding a minimum-cost MMDC between $A$ and $B$ using the well-known Hungarian algorithm.

\end{abstract}

\section{Introduction}
A \textit {matching} between two sets $A$ and $B$ defines a relationship between them. A \textit{many-to-many matching} between $A$ and $B$ maps each element of $A$ to at least one element of $B$ and vice-versa. A \textit{perfect matching} is a matching where each element is matched to a unique element. Eiter and Mannila \cite{Eiter} solved the many-to-many matching problem in $O(n^3)$ time by reducing it to the minimum weight \textit{perfect matching} problem in a bipartite graph, where $|A|+|B|=n$.  We refer the readers to \cite{Imanparast} for a comprehensive survey on the matching theory and algorithms.

Let $A=\{a_1,a_2,\dots,a_s\}$ and $B=\{b_1,b_2,\dots,b_t\}$ be two sets, a \textit{many-to-many matching with demands and capacities} (MMDC) matches each element $a_i \in A$ to $\alpha_i \leq d_i \leq {\alpha}'_i$ elements of $B$, and each element $b_j\in B$ to $\beta_j \leq d'_j \leq {\beta}'_j$ elements of $A$. MMDC problem is a specific case of the maximum weight degree-constrained subgraph problem in general graphs that in which for each vertex $v$ with degree $deg(v)$ we have $l(v)\leq deg(v)\leq u(v)$ ($l(v)$ and $u(v)$ denote integer bounds), and has the time complexity of $O(n^2 \min(m\log n,n^2))$ \cite{Gabow1983}. In this paper, we present an algorithm that computes a minimum-cost MMDC between $A$ and $B$ with $|A|+|B|=n$ in $O(n^6)$ time using the basic Hungarian algorithm. Also, our algorithm computes an MMDC between two sets of points in the plane in $O(n^4poly(\log n))$ time using the modified Hungarian algorithm proposed in \cite{Bandyapadhyay}. Moreover, in bipartite graphs with low-range edge weights and dense graphs, our algorithm runs faster than its worst time complexity \cite{LOPES201950}. In fact, our algorithm imposes upper and lower bounds on the number of elements that can be matched to each element in any version of the Hungarian algorithm.

\section {Preliminaries}
\label{Preliminaries}

Given an undirected bipartite graph $G=(A \cup B, E)$, a \textit {matching} in $G$ is a subset of the edges $M \subseteq E$, such that each vertex $v \in A \cup B$ is incident to at most one edge of $M$. Let $Weight(a,b)$ denote the weight of the edge $(a,b)$, the weight of the matching $M$ is the sum of the weights of all edges in $M$, hence $$Weight(M)=\sum_{e \in M}Weight(e).$$
A \textit {minimum weight matching} $M$ is a matching that for any other matching $M'$, we have $Weight(M') \geq Weight(M)$.

An \textit{alternating path} is a path with the edges alternating between $M$ and $E-M$. A \textit {matched vertex} is a vertex that is incident to an edge in $M$. A vertex that is not matched is a \textit{free vertex}. An alternating path whose both endpoints are free is an \textit {augmenting path}.

A \textit{vertex labeling} is a function $l: V \rightarrow \Bbb{R} $ that assigns a label to each vertex $v \in V$. A vertex labeling with $l(a)+l(b) \leq Weight(a,b)$ for all $a \in A$ and $b \in B$ is a \textit{feasible labeling}. The \textit {equality graph} of a feasible labeling $l$ is a graph $G=(V,E_l)$ with $E_l=\{(a,b)| l(a)+l(b)=Weight(a,b)\}$. The set of the \textit{neighbors} of a vertex $u \in V$ is defined as $N_l(u)=\{v| (v,u) \in E_l\}$. Given a set of the vertices $S \subseteq V$, the neighbors of $S$ is $N_l(S)=\bigcup_{u \in S} N_l(u)$.

\newtheorem{lemma}{Lemma}
\begin{lemma}
\label{lem1}
Let $G=(A\cup B, E)$ be an undirected bipartite graph, and $l$ be a feasible labeling of $G$. Let $S \subseteq A$ with $T=N_l(S)\neq B$. Assume that the labels of the vertices of $G$ are updated as follows:

 \begin{itemize}
   \item if $v\in S$, then $l'(v)=l(v)+\alpha_l$.
   \item if $v \in T$, then $l'(v)=l(v)-\alpha_l$.
   \item otherwise, $l'(v)=l(v)$
 \end{itemize}
that in which $$\alpha_l=\min_{a_i \in S, b_j \notin T}\{Weight(a_i,b_j)-l(a_i)-l(b_j)\}.$$
Then, $l'$ is also a feasible labeling with $E_l \subset E'_l$.
\end{lemma}

\textbf {Proof.} Note that $l$ is a feasible labeling, so we have \newline$l(a)+l(b)\leq Weight(a,b)$ for each edge $(a,b)$ of $E$. \newline After updating the labels, four cases arise:
\begin{itemize}
\item $a \in S$ and $b \in T$. In this case, we have $$l'(a)+l'(b)=l(a)+\alpha_l+l(b)-\alpha_l=l(a)+l(b)\leq Weight(a,b).$$
\item $a \notin S$ and $b \notin T$. Then, we have $$l'(a)+l'(b)=l(a)+l(b)\leq Weight(a,b).$$
\item $a \notin S$ and $b \in T$. We see that $$l'(a)+l'(b)=l(a)+l(b)-\alpha_l<l(a)+l(b)\leq Weight(a,b).$$
\item $a \in S$ and $b \notin T$. In this situation, we have
$$l'(a)+l'(b)=l(a)+\alpha_l+l(b).$$
Then, two cases arise:
\begin{itemize}

\item $Weight(a,b)-l(a)-l(b)=\alpha_l$. Then,
$$l'(a)+l'(b)=l(a)+\alpha_l+l(b)=l(a)-l(a)-l(b)+Weight(a,b)+l(b)=Weight(a,b).$$ Hence, $E_l \subset E_{l'}$.

\item $Weight(a,b)-l(a)-l(b)>\alpha_l$. Obviously
$$l'(a)+l'(b)=l(a)+\alpha_l+l(b)\leq Weight(a,b).$$

\end{itemize}

\end{itemize}
\qed

\newtheorem{theorem}{Theorem}
\begin{theorem}
Let $l$ be a feasible labeling and $M$ a perfect matching in $E_l$. Then, $M$ is a minimum weight matching \cite{Kuhn}.
\end{theorem}

In the following, we briefly describe the basic Hungarian algorithm which computes a minimum weight perfect matching in an undirected bipartite graph $G=(A \cup B, E)$ with $|A|=|B|=n$ (Algorithm \ref{BasicHungarian}) \cite{Kuhn,Munkres}.

\vspace{0.5cm}

\begin{algorithm}

\caption{The Basic Hungarian algorithm($G=(A\cup B,E)$)}
\label{BasicHungarian}

 Let $l(b_j)=0, \ for \ all \ 1 \le j \le n$\;
 $l(a_i)=\min_{j=1}^n Weight(a_i,b_j)\  for\  all \ 1 \le i \le n$\;
 $M= \emptyset$\;
\While {$M$ is not perfect}{

Select a free vertex $a_i  \in A$ and set $S = \{a_i\}$, $T=\emptyset$\;
\For{$j\gets 1, n$}{
 $slack[j]=l(a_i)+l(b_j)-Weight(a_i,b_j)$\;
}
  \Repeat{$u$ is free}{
     \If {$N_l(S)=T$}{
         $\alpha_l=\min_{b_j \notin T}slack[j]$\;
         $Update(l)$ 
\ForAll {$b_j \notin T$}{
  $slack[j]=slack[j]+\alpha_l$\;
}
        }
        Select $u  \in N_l (S)-T$\;
          \If {$u$ is not free}{
             $S = S \cup \{z\},T = T \cup \{u\}$\;
\For{$j\gets 1, n$}{
  $slack[j]=\min (Weight(z,b_j)-l(z)-l(b_j),slack[j])$\;
}

          }
       }  
   $Augment(M)$\;
}
\Return $M$\;

 \end{algorithm}

In Lines 1--2, we label all vertices of $B$ with zero and each vertex $a_i \in A$ with $\min_{j=1}^n Weight(a_i,b_j)$ to get an initial feasible labeling. Note that $M$ can be empty (Line 3). In each iteration of the while loop of Lines 4--20, two free vertices $a_i$ and $b_j$ are matched, so it iterates $O(n)$ times. Using the array $slack[1,\dots,n]$, we can run each iteration of this loop in $O(n^2)$ time. The repeat loop runs at most $O(n)$ times until finding a free vertex $b_j$. In Line 10, we can compute the value of $\alpha_l$ by:
$$\alpha_l=\min_{b_j \notin T}slack[j],$$
in $O(n)$ time. After computing $\alpha_l$ and updating the labels of the vertices, we must also update the values of the slacks. This can be done using:
$$for \ all\  b_j \notin T, slack[j]=slack[j]+\alpha_l.$$ In Line 11, we update the feasible labeling $l$ such that $N_l(S) \neq T$. In Line 15 of Algorithm \ref{BasicHungarian}, when a vertex is moved from $\bar S$ to $S$, the values of $slack[1,\dots,n]$ must be updated. This is done in $O(n)$ time. $O(n)$ vertices are moved from $\bar S$ to $S$, so it takes the total time of $O(n^2)$.

The value of $\alpha_l$ may be computed at most $O(n)$ times in $O(n)$ time, so running each iteration takes at most $O(n^2)$ time. So, the time complexity of the basic Hungarian algorithm is $O(n^3)$. The Hungarian algorithm in the worst case does the repeat loop of the algorithm in $O(n^2)$ overall time; updating the labels using the function $Update(l)$ adds only one more edge to $l'$ ($|E_{l'}|=|E_l|+1$), but we observe that in practice, in bipartite graphs with low-range edge weights and dense graphs, more edges are added to $E_l$ after running $Update(l)$ \cite{LOPES201950}.

\section {MATCHING ALGORITHM}
\label{Matchingsec}

We construct a complete bipartite graph such that by applying the Hungarian method on it, the demands and capacity limitations of the elements are satisfied. In the following, we explain how our complete bipartite graph $G$ is constructed.

We represent a set of the related vertices using a rectangle, each connection between two vertices with a line and each vertex with a circle. So, a connection between two vertices is shown using a line that connects the two corresponding circles. A \textit{complete connection} between two sets is a connection where each vertex of one set is connected to all vertices of the other set. We show a complete connection using a line connecting the two corresponding sets.

Let $S \cup T$ be a bipartition of $G$, where $S=(\bigcup_{i=1}^{s} A_i) \cup (\bigcup_{i=1}^{s}{A'}_i ) \cup ( \bigcup_{j=1}^{t} X_j) \cup ( \bigcup_{j=1}^{t} W_j)$ and $T=\bigcup_{i=1}^{s} Bset_i$. The vertices of the sets $A_i$, $Bset_i$, and ${A'}_i$ for all $1\leq i\leq s$ are called the \textit{main vertices}, since they are copies of the input elements. On the other hand, the vertices of the sets $X_j$ and $W_j$ for all $1\leq j\leq t$ are called the \textit{dummy vertices}. All edges $(a,b)$ that their both end vertices are main vertices, that is $a \in A_i \cup {A}'_i$ and $b \in Bset_i$ for $1\leq i\leq s$, are called the \textit{main edges}.

The Hungarian method computes a \textit {perfect} matching where each vertex is incident to a unique edge. We aim to find an MMDC matching in which two or more vertices may be mapped to the same vertex, that is a vertex may be selected more than once. So, our constructed graph contains multiple copies of each element to simulate this situation. Let $A_i=\{a_{i1},\dots,a_{i\alpha_i }\}$ for $1\leq i\leq s$ be the set of the $\alpha_i$ copies of the element $a_i$. Each set $A_i$ is completely connected to the set $Bset_i= \{b_{1i},\dots,b_{ti}\}$ for $1\leq i\leq s$. This complete connection is shown using a line connecting the two corresponding rectangles of $A_i$ and $Bset_i$. Note that $Weight(a_{ik},b_{ji})=\delta(a_i,b_j)$, where $\delta(a_i,b_j)$ is the cost of matching $a_i$ to $b_j$. Each $A_i$ set guarantees that each element $a_i \in A$ is matched to at least $\alpha_i$ elements of $B$.

Note that each vertex of $A$ has a limited capacity, i.e. it must be matched to at most a given number of the elements of the other set. Each element $a_i$ is copied $({\alpha}'i-\alpha_i)$ times and constitutes the ${A'}_i$ set. Let ${A'}_i=\{{a'}_{i1},\dots,{a'}_{i({\alpha}'i-\alpha_i)}\}$ for $1\leq i\leq s$. ${A'}_i$ sets guarantee that each element $b_j \in B$ is matched to at least $\beta_j$ elements of $A$. Moreover, each set ${A'}_i$ assures that each element $a_i$ is matched to at most ${\alpha}'_i$ elements. Each ${A'}_i$ is completely connected to $Bset_i$, where $Weight({a'}_{id},b_{ji})$ is equal to $\delta(a_i,b_j)$ for all $1 \leq d \leq ({\alpha}'i-\alpha_i)$.

Assume that all vertices $b_{ji}$ for $1\leq i\leq s$ constitute sets, called $B_j$. In fact, the set $B_j$ is $s$ copies of $b_j$. We use the $W_j= \{w_{j1},\dots,w_{j(s-{\beta '}_j)}\}$ set to limit the number of the elements that can be matched to $b_j \in B$ for $1 \leq j \leq t$. There is a zero weighted complete connection between the vertices of $B_j$ and $W_j$ for $1\leq j\leq t$.

Let $X_j=\{x_{j1},\dots,x_{j({\beta}'_j-\beta_j)}\}$ and $\gamma=\max(\delta(a_i,b_j))$ for all $1\leq i\leq s$ and $1 \leq j \leq t$. Select an arbitrary number ${\gamma}'$ such that ${\gamma}'> \gamma$, there exists a ${\gamma}'$ weighted complete connection between the vertices of $B_j$ and $X_j$ for all $1\leq j \leq t$. $X_j$ sets guarantee that the matching is a minimum-cost matching. Note that the priority of the set $Bset_j$ is the sets $A_i$ and $A'_i$.

There exists another set that compensates the bipartite graph, called $Y$. The input of the Hungarian algorithm is a complete bipartite graph, i.e. both parts of the input bipartite graph have an equal number of vertices. Therefore, we should balance two parts of our constructed bipartite graph before using the Hungarian algorithm.
We have $$|S|=|\bigcup_{i=1}^{s} A_i| + |\bigcup_{i=1}^{s}{A'}_i|+|\bigcup_{j=1}^{t} X_j|+|\bigcup_{j=1}^{t} W_j|$$
$$=\sum_{i=1}^s\alpha_i+\sum_{i=1}^s{\alpha}'_i-\sum_{i=1}^s\alpha_i+\sum_{j=1}^t{\beta '}_j-\sum_{j=1}^t\beta_j+s*t-\sum_{j=1}^t{\beta '}_j$$$$=\sum_{i=1}^s{\alpha}'_i+(s*t)-\sum_{j=1}^t\beta_j,$$
and $$|T|=|\bigcup_{i=1}^{s} Bset_i|=(s*t).$$

\begin{figure}
\vspace{0cm}
\hspace{0cm}
\resizebox{1\textwidth}{!}{%
  \includegraphics{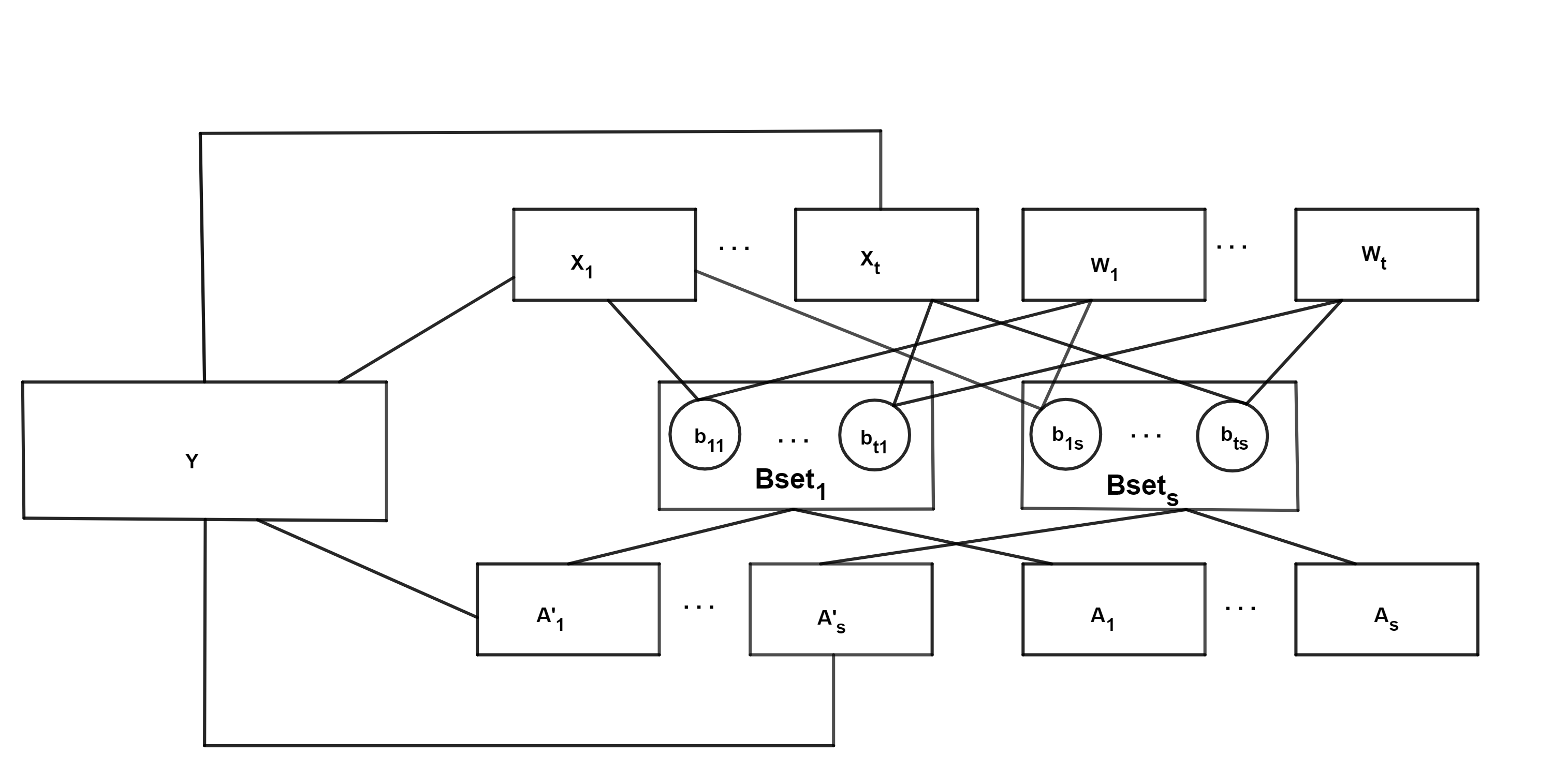}
}
\vspace{-0.7cm}
\caption{The constructed complete bipartite graph $G$ by our algorithm.}
\label{fig:1}       
\end{figure}

Let $|Y|=\sum_ {i=1}^s\alpha '_i-\sum_{j=1}^t\beta_j$. The compensator set $Y$ is inserted to $T$ as follows. Note that we have $\sum_ {i=1}^s\alpha '_i>\sum_{j=1}^t\beta_j$. There is a complete connection between $X_j$ and $Y$ that in which the weight of the edges is an arbitrary number ${\gamma}''$ with $ {\gamma}' <{\gamma}''$. Consequently, the priority of the vertices of $X_j$ is the vertices of $B_j$ set. Moreover, ${A'}_i$ is completely connected to $Y$ with $\gamma'$ weighted edges. Our constructed complete bipartite graph $G$ is shown in Figure \ref{fig:1}.

We claim that from a minimum weight perfect matching in $G=S \cup T$ denoted by $M$, we can get a minimum-cost MMDC between $A$ and $B$. Let $Mian(M)$ be the union of the main edges of the minimum weight perfect matching $M$ in $G$. In the following, we prove that the weight of $Mian(M)$ is equal to the cost of a minimum-cost MMDC between $A$ and $B$, called $L$. Let $c(L)$ denote the cost of $L$.

\begin{lemma}
\label{lem11}
$Weight(Main(M))\leq c(L)$.
\end{lemma}
\textbf {Proof.} We get from $L$ a perfect matching $M'$ in our complete bipartite graph $G$, such that the weight of the union of the main edges of $M'$, $Main(M')$, is equal to the cost of $L$, that is $Weight(Mian(M'))=c(L)$.

Let $p_i$ be the number of the elements $b_j \in B$ that are matched to $a_i\in A$ in $L$. It is obvious that $\alpha_i\leq p_i \leq {\alpha '}_i$. Firstly, for each pairing $(a_i,b_j)$ in $L$, we connect $b_{ji}$ to one of the unmatched vertices of $A_i$, that is $a_{ik}$ with $1\leq k \leq \alpha_i$, until there does not exist any unmatched vertex in $A_i$. Then, depending on the value of $p_i$ two cases arise:
\begin{itemize}
\item either $p_i=\alpha_i$. In this situation, we add the $\gamma'$ weighted edges of $G$ connecting each ${a'}_{ij}\in {A'}_i$ to one of the unmatched vertices of $Y$ for all $1\leq j \leq ({\alpha}'_i-\alpha_i)$.

\item or $p_i>\alpha_i$. In this case, we need to match $p_i-\alpha_i$ number of the vertices of ${A'}_i$ with the vertices of $Bset_i$. So, for each pairing $(a_i,b_j)$ of the $p_i-\alpha_i$ remaining pairings, we add an edge of $G$ connecting ${a'}_{ij}$ to $b_{ji}$. Then, if yet there exist other vertices of ${A'}_i$ that have not been matched to any vertex (i.e. $p_i<\alpha'_i$); for each of them, we select an edge of $G$ connecting it to an unmatched vertex of $Y$, and add it to $M'$.

\end{itemize}

Then, for each $w_{jk} \in W_j$ we add the edge of $G$ that connects it to an unmatched vertex of $B_j$. The vertices of $X_j$ are matched to the vertices of $B_j$, unless no vertices remain unmatched in $B_j$. So, we first add the edges connecting the vertices of $X_j$ to the remaining unmatched vertices of $B_j$. Then, we add the edges connecting the unmatched vertices of $X_j$, if exist, to the unmatched vertices of $Y$.

Since all vertices of $G$ are selected once, $M'$ is a perfect matching. For each $(a_i,b_j)\in L$, there is an edge with equal weight in $Mian(M')$, so $Weight(Mian(M'))=c(L)$.\qed

\begin{lemma}
\label{lem13}
Let $M$ be a minimum weight perfect matching in $G$. Then, for any perfect matching in $G$ denoted by $M'$ we have
$$Weight(Main(M))\leq Weight(Main(M')).$$
\end{lemma}

\textbf {Proof.} Observe that we have:
$$Weight(M)=Weight(Main(M))+Weight(M-Main(M)).$$

Note that for a minimum-cost MMDC between $A$ and $B$ denoted by $L$ we have $$|L|=\max(\sum_ {i=1}^s\alpha_i,\sum_{j=1}^t\beta_j).$$
Given any perfect matching $M''$ in $G$, the set $M''-Main(M'')$ contains:
\begin{itemize}
  \item the zero weighted edges connecting the vertices of $W_j$ to the vertices of $B_j$ for $1\leq j\leq t$, with the total number of $s*t-\sum_{j=1}^t{\beta '}_j$,
  \item the $\gamma'$ weighted edges connecting $\sum_{j=1}^t\beta'_j-|L|$ number of the vertices of $X_j$ to the vertices of $B_j$ for $1\leq j\leq t$,
  \item the $\gamma''$ weighted edges connecting $|L|-\sum_{j=1}^t\beta_j$ number of the vertex of $X_j$ to $Y$ for $1\leq j\leq t$,
  \item the $\gamma'$ weighted edges connecting $\sum_{i=1}^s\alpha'_i-|L|$ number of the vertex of $A'_i$ to $Y$ for $1\leq i\leq s$.
\end{itemize}

Thus

\begin{equation}
\nonumber
\begin{split}
Weight(M''-Main(M''))=(\sum_{j=1}^t\beta'_j-|L|)*\gamma'\\
+(|L|-\sum_{j=1}^t\beta_j)*\gamma''+(\sum_{i=1}^s\alpha'_i-|L|)*\gamma'.&
\end{split}
\end{equation}

So, we have:
$$Weight(M-Main(M))=Weight(M'-Main(M')).$$

Note that $M$ is a minimum weight perfect matching in $G$, thus
$$Weight(M)\leq Weight(M'),$$
and so

\begin{equation}
\nonumber
\begin{split}
Weight(M-Main(M))+Weight(Main(M))\leq\\
Weight(M'-Main(M'))+Weight(Main(M')),&
\end{split}
\end{equation}


Therefore, we have:
$$Weight(Main(M))\leq Weight(Main(M')).$$

Note that $Weight(Main(M'))=c(L)$,
so
$$Weight(Main(M))\leq c(L).$$

\qed

\begin{lemma}
\label{lem12}
$c(L)\leq Weight(Main(M))$.
\end{lemma}

\textbf {Proof.} From $Main(M)$, we get an MMDC between $A$ and $B$ denoted by $L'$, such that the cost of $L'$ is equal to the weight of $Main(M)$, that is $Weight(Main(M))=c(L')$.
For each edge $m \in M$, if $m=(a_{ik},b_{ji} )$ or $m=({a'}_{ik},b_{ji} )$, we add the pairing $(a_i,b_j )$ to $L'$. Otherwise, no pairing is added to $L'$.

For each $a_i\in A$ for $1\leq i\leq s$, there exists the set $A_i$ in $G$ with $\alpha_i$ vertices which are connected only to one set, $Bset_i$. So, the vertices of each $A_i$ are matched to some vertices of $Bset_i$, that is $b_{ji}$ for $1\leq j\leq t$. Hence, each $a_i\in A$ for $1\leq i\leq s$ is matched to at least $\alpha_i$ elements of $B$, and the demand of $a_i$ is satisfied. In $G$ there exist $\alpha_i$ plus ${\alpha}'_i-\alpha_i$ copies of each element $a_i$, that is the vertices of $A_i$ plus the vertices of $A'_i$. So, the number of elements that are matched to each $a_i\in A$ is at most ${\alpha}'_i$.

Consider the sets $B_j$ with $1\leq j\leq t$, recall that $B_j=\{b_{ji}|1\leq i\leq s\}$ and the vertices of $W_j$ are connected to $B_j$ for $1\leq j\leq t$ by zero weighted edges. $W_j$ is connected only to $B_j$, so the vertices of $W_j$ are matched to $s-{\beta '}_j$ number of the vertices of $B_j$, and ${\beta '}_j$ number of the vertices remains unmatched in $B_j$. Suppose that $k$ vertices of ${\beta '}_j$ vertices in $B_j$ are matched to the vertices of $A_i$ sets for $1\leq i\leq s$, so the ${\beta '}_j-k$ remaining vertices of $B_j$ should be matched to the other sets that are connected to it. We discuss two cases, depending on the value of $k$.
\begin{itemize}
\item if $k<\beta_j$ then $({\beta '}_j-k)>({\beta '}_j-\beta_j)$. Then, $X_j$ selects the ${\beta '}_j-\beta_j$ vertices of the remaining vertices of $B_j$. we have
$$({\beta '}_j-k)-({\beta '}_j-\beta_j)={\beta '}_j-k-{\beta '}_j+\beta_j=\beta_j-k>0,$$ so the remaining $\beta_j-k$ vertices of $B_j$ are matched to the vertices of ${A'}_i$ sets. Note that $k$ vertices of the vertices $b_{ji}$ for all $1\leq i\leq s$ are matched to the vertices of $A_i$ sets and $\beta_j-k$ vertices of them are matched to ${A'}_i$ sets. The demand of the element $b_j$ is satisfied, since $$\beta_j-k+k=\beta_j.$$
\item if $k>\beta_j$ then $({\beta '}_j-k)<({\beta '}_j-\beta_j)$ and all the $({\beta '}_j-k)$ remaining members of $B_j$ are matched to the vertices of $X_j$.
\end{itemize}

The cost of $L'$ is equal to the weight of $Main(M)$, i.e. $c(L')=Weight(Main(M))$, since for each edge of $Main(M)$, we add a pairing with equal cost to $L'$. On the other hand, $L'$ is an MMDC between $A$ and $B$. $L$ is a minimum-cost MMDC between $A$ and $B$, so $c(L)\leq c(L')$. Thus $$c(L)\leq Weight(Main(M)).$$\qed

\begin{theorem}
\label{the}
Let $M$ be a minimum weight perfect matching in $G$, and let $L$ be a minimum-cost MMDC between $A$ and $B$. Then, $Weight(Main(M))=c(L)$.
\end{theorem}
\textbf {Proof.} By Lemma \ref{lem11} and Lemma \ref{lem12}, we have $Weight(Main(M))\leq c(L)$ and $Weight(Main(M))\geq c(L)$, respectively. So we have $Weight(Main(M))=c(L)$.\qed

Recall that the time complexity of the Basic Hungarian algorithm is $O(n^3 )$, where the number of the vertices of the input graph is $O(n)$. The number of the vertices of our complete bipartite graph is $O(n^2)$, so the complexity of our algorithm is $O(n^6)$, but in bipartite graphs with low-range edge weights and dense graphs, our algorithm runs well \cite{LOPES201950}.

\section{Conclusion}
\label{ConclusionSect}
We presented an algorithm for the minimum-cost MMDC problem by advantage of the Hungarian algorithm. It is expected that the complexity of the MMDC problem will be reduced by exploiting the geometric information; the one and two dimensional versions of this problem remain open.

\section*{Data Availability}
No data were used to support this study.

\section*{Conflicts of Interest}
There is no conﬂict of interest to declare.

\end{document}